\documentclass[aps,prd,10pt,twocolumn,nofootinbib]{revtex4}

\usepackage{epsfig}
\usepackage{bm}
\usepackage{latexsym}
\usepackage{natbib}
\usepackage{url}
\usepackage{dcolumn}
\usepackage{color}
\usepackage{amsfonts,amssymb,amsmath}
\usepackage{graphicx,epsfig}
\usepackage{psfrag}
\usepackage{subfigure}
\usepackage{hyperref}
\hypersetup{colorlinks=true}
\usepackage{mathtools}
\usepackage{enumitem}
\usepackage{float}
\usepackage[dvipsnames]{xcolor}
\usepackage{lipsum}

\newcommand{\beq}{\begin{equation}}
\newcommand{\eeq}{\end{equation}}
\newcommand{\bea}{\begin{eqnarray}}
\newcommand{\eea}{\end{eqnarray}}
\newcommand{\beas}{\begin{eqnarray*}}
\newcommand{\eeas}{\end{eqnarray*}}

\newcommand{\bquo}{\begin{quote}}
\newcommand{\equo}{\end{quote}}
\renewcommand{\(}{\begin{equation}}
\renewcommand{\)}{\end{equation}}

\def\IZ{{\mathbb Z}}
\def\IR{{\mathbb R}}

\def\Tr{ \hbox{\rm Tr}}

\def\exo{\hbox {\rm exp}}

\def\dag{{}^{\dagger}}

\def\l({\left(}\def\r){\right)}

\def\IZ{{\mathbb Z}}
\def\IR{{\mathbb R}}

\newcommand\Mycomb[2][^n]{\prescript{#1\mkern-0.5mu}{}C_{#2}}

\def\beq{\begin{equation}}
\def\eeq{\end{equation}}

\begin{document}

\title{A Large-$N$ Phase Transition in a Finite Lattice Gauge Theory}

\author{
Nirmalya Brahma$^{a,}$\footnote{nirmalyab@iisc.ac.in}, and Chethan Krishnan$^{a,}$\footnote{chethan.krishnan@gmail.com}
}

\affiliation{
$^a$ Center for High Energy Physics,
Indian Institute of Science, Bangalore 560012, India
}

\begin{abstract}
We consider gauge theories of non-Abelian $finite$ groups, and discuss the 1+1 dimensional lattice gauge theory of the permutation group $S_N$ as an illustrative example. The partition function at finite $N$ can be written explicitly in a compact form using properties of $S_N$ conjugacy classes. A natural large-$N$ limit exists with a new 't Hooft coupling, $\lambda=g^2 \log N$. We identify a Gross-Witten-Wadia-like phase transition at infinite $N$, at $\lambda=2$. It is first order. An analogue of the string tension can be computed from the Wilson loop expectation value, and it jumps from zero to a finite value. We view this as a type of large-$N$ (de-)confinement transition. Our holographic motivations for considering such theories are briefly discussed.


%
 
\end{abstract}

\maketitle

\section{Introduction and Conclusion}

In this paper, we will discuss lattice gauge theory of the finite group $S_N$ with gauge coupling $g$ in two dimensions, and consider its large-$N$ limit. We will see that there is a natural 't Hooft coupling $\lambda \equiv g^2 \log N$, and that there is a first order phase transition as $\lambda$ changes from small to large values at infinite $N$. Even though the string tension is not a directly meaningful quantity to define in these theories, we will find that there is a closely related quantity one {\em can} define via the Wilson loop expectation value. This ``string tension" undergoes a jump from zero to a finite value at the transition, suggesting that it may be useful to view this as an analogue of a large-$N$ (de-)confinement transition. 

Our discussion is closely parallel to the well known Gross-Witten-Wadia transition \cite{GW, Wadia1, Wadia2} in $SU(N)$ lattice Yang-Mills theory in 1+1 dimensions where the relevant 't Hooft coupling is $\lambda_{GWW} = g^2_{YM} N$, the transition is third order, and we are always in the confining phase. Our presentation will be closely modeled on the discussion in these papers. 

One of our motivations for considering theories with finite gauge groups is the possibility that they may provide a simple setting for the introduction of boundary gauge invariance in discussions of holographic tensor networks and quantum error correcting codes. In the currently existing discussions of these ideas (see eg., \cite{Swingle, Harlow}), the gauge invariance of the boundary theory does not play any role at all. On the other hand, in all known explicit examples of the holographic correspondence (see eg. \cite{Kiriakos}), the dual theory has some form of gauge invariance. So it seems to us that {\em boundary} gauge invariance may have a more substantive message  about questions of {\em bulk} locality, than hitherto appreciated\footnote{Some thought-provoking, but ultimately inconclusive previous discussions can be found in \cite{Polchinski, Yang, Kabir}.}. Much of the work on tensor network quantum codes are in the language of discrete toy models, so it is natural to consider discrete toy models for boundary gauge invariance as a  starting point. Since we would like to have an interesting large-$N$ limit, {\it non-Abelian} finite groups become a natural option.  

In this paper, we will however not say much about our holographic goals (but see \cite{future}). 
Our present interst in these theories stems from the fact that in the $S_N$ example we consider, (a) the finite-$N$ partition function is explicitly calculable and can be written in a simple enough form, (b) the analogue of a large-$N$ first order (de-)confinement transition is present, and (c) the 't Hooft coupling has an interesting new scaling. More reasons for interest in these theories will be discussed as they emerge, and also in the concluding section.

Finite gauge groups have been explored on the lattice ever since Wegner's work on $\IZ_2$ gauge theory \cite{Wegner}, see \cite{Kogut} for a pedagogical review of related ideas. One discussion of finite groups on lattices that we were influenced by, is \cite{Trivedi}, which also contains some previous references. The group $S_N$ has recently played an interesting role in a loosely related context in \cite{Hanada}. 

 
\section{1+1 D Finite Lattice Gauge Theory}

In this section, we will review our general setup to study lattice gauge theories with a gauge group $G$ at each link. Our discussion is a direct adaptation of \cite{GW} to general groups $G$, and our primary goal is to set up the notation. We will be interested in the case where $G$ is a finite group, and we will refer to such theories as {\it finite lattice gauge theories}. We will emphasize some points where the distinctions are worth noting. 

Consider a 1+1 dimensional lattice with lattice spacing $a$ (along both temporal and spatial directions). Denote the lattice vector in the temporal direction by $\hat{i}_{0}$ and that in the spatial direction by $\hat{i}_{1}$. Using these unit vectors we can label each site as $\vec{n}(=n_{0}\hat{i}_{0}+n_{1}\hat{i}_{1})$. The link starting at the site $\vec{n}$ and ending at $\vec{n}+\vec{\mu}$ is labelled by $\left(\vec{n},\vec{\mu}\right)$, where $\vec{\mu} \in \{\hat{i}_{0},\hat{i}_{1}\}$. The dynamical variable associated with each link is a unitary matrix $U\left(\vec{n},\vec{\mu}\right)$
where $U\left(\vec{n},\vec{\mu}\right)\in R_G$, where we will take $R_G$ to be a unitary (typically irreducible) representation of $G$.  A particular link $\left(\vec{n},\vec{\mu}\right)$ can also be labelled as $\left(\vec{n}+\vec{\mu},-\vec{\mu}\right)$, from which we can write,
 \begin{equation}
 \label{real}
 U\left(\vec{n}+\vec{\mu},-\vec{\mu}\right)=U^{\dag}\left(\vec{n},\vec{\mu}\right)
\end{equation} 
Under local gauge transformations, the link variables transform as:
\begin{equation}
U(\vec{n},\vec{\mu})\rightarrow V(\vec{n})U(\vec{n},\vec{\mu})V^{\dag}(\vec{n}+\vec{\mu})
\end{equation}
for arbitrary $V(\vec{n})$ sitting at each site $\vec{n}$ with the condition that $V(\vec{n})V^{\dag}(\vec{n})=V(\vec{n})^{\dag}V(\vec{n})=I$.
Then the gauge-invariant Wilson action in these variables can be written as:
\begin{widetext}
\vspace{-0.29in}
\bea
S&=&\frac{1}{g^{2}}\sum_{\vec{n}}\Tr\left[U(\vec{n},\hat{i}_{1})U(\vec{n}+\hat{i}_{1},\hat{i}_{0})U(\vec{n}+\hat{i}_{1}+\hat{i}_{0},-\hat{i}_{1})U(\vec{n}+\hat{i}_{0},-\hat{i}_{0})+\text{H.c.}\right] \nonumber \\
&=&\frac{1}{g^{2}}\sum_{\vec{n}}\Tr\left[U(\vec{n},\hat{i}_{1})U(\vec{n}+\hat{i}_{1},\hat{i}_{0})U^{\dag}(\vec{n}+\hat{i}_{0},\hat{i}_{1})U^{\dag}(\vec{n},\hat{i}_{0})+\text{H.c.}\right]
\eea
\end{widetext}
where in the second equality we used (\ref{real}). In words, the term inside the trace is a product of $R_G$-matrix variables around a single plaquette (smallest cell) of the lattice while the sum adds the contribution from all the plaquettes.

The gauge-invariance of the action presents us the possibility of making a gauge choice. Choosing the ``temporal" gauge \cite{GW} 
\begin{equation}
\label{gauge choice}
U(\vec{n},\hat{i}_{0})=I \quad \forall \vec{n}
\end{equation}
the action becomes
\begin{equation}
S=\frac{1}{g^{2}}\sum_{\vec{n}}\left[U(\vec{n},\hat{i}_{1})U^{\dag}(\vec{n}+\hat{i}_{0},\hat{i}_{1})+\text{H.c.}\right]
\end{equation}
By a change of variables
\begin{equation}
\label{change variable}
U(\vec{n}+\hat{i}_{0},\hat{i}_{1})\equiv W(\vec{n})U(\vec{n},\hat{i}_{1})   
\end{equation}
where again $W(\vec{n})\in R_G$, the action further simplifies to:
\begin{equation}
\label{action}
S=\frac{1}{g^{2}}\sum_{\vec{n}}\Tr\left[W(\vec{n})+W^{\dag}(\vec{n})\right]
\end{equation}

In this paper, we will concern ourselves with \textit{finite} gauge groups. Using the form of the action in (\ref{action}) the partition function for finite groups can be written as
\begin{align}
Z&=\frac{1}{|G|^P}\sum_{W(\vec{n}) \in G}\exo{\left(\frac{1}{g^{2}}\sum_{\vec{n}}\Tr\left[W(\vec{n})+W^{\dag}(\vec{n})\right]\right)}\\
&=(z)^{P},
\end{align}
where $P (=V/a^{2})$ is the total number of plaquettes, $|G| $ is the order of the group $G$, $V$ is the volume (essentially area) of our two-dimensional lattice, and $z$ is the partition function for a single plaquette given by
\(
\label{partition function}
z=\frac{1}{|G|}\sum_{W \in G}\exo{\left(\frac{1}{g^{2}}\Tr\left[W+W^{\dag}\right]\right)}.
\)
The normalizations are chosen so that when the summand is trivial (identity) the sum is unity. Note that this is just the Haar measure for finite groups. We see that the evaluation of the partition function for the whole system reduces to the computation of the partition function for a single plaquette. This 
is a characteristic of two-dimensional gauge theories defined on $\IR^2$ with open boundary conditions. See \cite{Wadia1} for a discussion on the cylinder.  

The Wilson loop operator is defined by \cite{GW}:
\(
\hat{W}_{L}=\frac{1}{{\rm dim} R_G}\Tr\left[\prod_{L}U\right] \label{Wilson}
\)
where the product is over a closed loop $L$. 
Consider a rectangular loop with length along time-axis and space-axis be $Ta$ and $Ra$, respectively. Using (\ref{gauge choice}), the Wilson loop operator $\hat{W}_{L}$ becomes 
\bea
\frac{1}{{\rm dim} R_G}\Tr\left[\prod_{k=0}^{R-1}U(T\hat{i}_{0}+k\hat{i}_{1},\hat{i}_{1}) \prod_{k=R-1}^{0}U^{\dag}(0\hat{i}_{0}+k\hat{i}_{1},\hat{i}_{1})\right]. \nonumber 
\eea
With a further gauge choice of $U(0\hat{i}_{0}+k\hat{i}_{1},\hat{i}_{1})=I$ $\forall k$,
the only non-trivial links of the loop are the space-like links at time step $T$. Consequently, 
\begin{align}
\hat{W}_{L}&=\frac{1}{{\rm dim} R_G}\Tr\left[\prod_{k=0}^{R-1}U(T\hat{i}_{0}+k\hat{i}_{1},\hat{i}_{1})\right], \nonumber\\
&=\frac{1}{{\rm dim} R_G}\Tr\left[\prod_{k=0}^{R-1}\prod_{j=T-1}^{0} W(j\hat{i}_{0}+k\hat{i}_{1})\right]
\end{align}
where in the last equality, we repeatedly used (\ref{change variable}) till we get $U(0\hat{i}_{0}+k\hat{i}_{1},\hat{i}_{1})$'s. Then the expectation value of the Wilson loop can be written as
\begin{widetext}
\bea
\label{wilson loop}
W_{L}(g^{2},N)=\left\langle\hat{W}_{L} \right\rangle
=\frac{1}{Z}\frac{1}{|G|^P}\sum_{W(\vec{n}) \in G}\exo{\left(\frac{1}{g^{2}}\sum_{\vec{n}}\Tr\left[W(\vec{n})+W^{\dag}(\vec{n})\right]\right)}\hat{W}_{L}
\eea
\end{widetext}
In the subsequent sections our discussions will be centered around the permutation group, $G=S_{N}$. Various nice properties of this group will enable us to obtain an exact, analytical expression for the partition function which will prove to be very convenient for our investigations on phase transition.

\section{Gauging the Permutation Group}

The permutation group $S_{N}$ is a finite, discrete group whose elements are given by permutations of $N$ objects. Evidently, the order of the group is $N!$. We will also use the defining representation of $N\times N$ permutation matrices as our representation $R_{S_N}$. In other words,  in the rest of this paper
\bea
|G|=N!, \ \ {\rm dim} R_G = N.
\eea
The permutation matrices are real and unitary. Let us note one minor point. Permutation matrices are a {\it reducible} representation of the permutation group, which splits into two irreps -- the trivial 1-dimensional representation and an $(N-1)$-dimensional complement \cite{Wikipedia}. Since the distinction will not be significant in the large-$N$ limit, we will present the discussion here in terms of the reducible permutation matrix representation. We have checked that working with the $(N-1)$-irrep only results in an extra subleading-in-$N$ overall factor of $\exp(2/g^2)$ in the partition function, which does not affect our claims.    

The trace (or the character of the matrix representation) in the action is equal to the sum of eigenvalues of the corresponding matrix.  This sum of eigenvalues is fixed by the conjugacy class, and does not depend on the permutations that preserve the conjugacy class. But this correspondence is non-unique, in the sense that more than one conjugacy classes might have the same value for the sum. The conjugacy classes are fixed by the cycle structures (see eg., \cite{Georgi}). If we take $k_{i}$ to be the number of cycles of length $i$ 
in a conjugacy class (with the implicit constraint that $\sum_{i=1}^{N}k_{i}i=N$), then the number of elements in a conjugacy class defined by a particular set of pairs $\{k_{i},i\}$, is given by 
\begin{equation}
    \frac{N!}{\prod_{i=1}^{N}i^{k_{i}}k_{i}!}.
\end{equation}

An $i$-cycle in the conjugacy class is reflected in the eigenvalue structure of the permutation matrix as the presence of $i$-the roots of unity. 
Hence the action of our theory, which contains a sum of eigenvalues, gets contributions only from the number of $1$-cycles in the conjugacy class, $k_{1}$. Putting all of this together into (\ref{partition function}), we can write the partition function for the group $S_{N}$ as 
\vspace{0.2in}
\(
\label{cojugacy partition}
z=\sum_{k_{1}}\sum_{k_{2}}\ldots\sum_{k_{N}}\frac{N!}{\prod_{i=1}^{N}i^{k_{i}}k_{i}!}\exo\left({\frac{2}{g^{2}}k_{1}}\right)
\)
\\
with the constraint $\sum_{i=1}^{N}k_{i}i=N$. The coefficient of the term with $2 k_{1}/g^2$ in the exponent is the sum of all conjugacy classes with that value of $k_{1}$.

We can obtain a simpler expression for the partition function by approaching the problem from a different viewpoint. Any row of a permutation matrix is composed of only one $1$ and rest of them are zero. The rows with $1$ as the diagonal element correspond to the 1-cycles. For a matrix to have $k_{1}$ 1-cycles, it must have $1$ as the diagonal elements in $k_{1}$ of the rows and $0$ as the diagonal element in $N-k_{1}$ of the other rows. We can choose these $N-k_{1}$ rows in $\Mycomb[N]{N-k_{1}}$ ways. In each of these rows, the $1$ can't go to the places where there is already a $1$ in other rows. That leaves us with $N-k_{1}$ choices. We have to find a place for the $N-k_{1}$ $1$'s among the $N-k_{1}$ slots such that none of the $1$'s land on the diagonal slot. This is equal to the {\it number of derangements} of a set of size $N-k_{1}$, which is equal to $!(N-k_{1})$, sometimes called the {\em subfactorial} \cite{subfactorial} of $(N-k_1)$:
\bea
!n \equiv n! \sum_{k=0}^n \frac{(-1)^k}{k!}
\eea
 Hence, we can write the total number of elements with $k_{1}$ $1$-cycles as
\begin{equation*}
   \frac{N!}{k_{1}!(N-k_{1})!}\times !(N-k_{1})
\end{equation*}
Then the partition function of (\ref{cojugacy partition}) becomes
\(
\label{subfactorial partition}
z=\sum_{k_{1}} !(N-k_{1})\frac{N!}{k_{1}!(N-k_{1})!}\exo\left({\frac{2}{g^{2}}k_{1}}\right)
\)
which implicitly shows that
\(
\sum_{k_{2}}\ldots\sum_{k_{N}}\frac{N!}{\prod_{i=2}^{N}i^{k_{i}}k_{i}!}=!(N-k_{1}) \frac{N!}{k_{1}!(N-k_{1})!} \label{identity}
\)
where the sum in the L.H.S is constrained by $\sum_{i=2}^{N}k_{i}i=N-k_{1}$. The identity \eqref{identity} can be easily checked on Mathematica for various cases.

From the properties of permutation matrices, we can also compute a nice expression for the Wilson loop to the leading order in $1/N$. Following \cite{GW} let us start by writing the sum for a particular $W(\vec{n})$ in the Wilson loop from (\ref{wilson loop}) as
\begin{widetext}
\bea
\frac{1}{N!}\sum_{W(\vec{n}) \in S_{N}}\exo{\left(\frac{2}{g^{2}}\Tr\left[W(\vec{n})\right]\right)}\frac{1}{N}\Tr\left[A W(\vec{n})B\right] 
=\frac{1}{N!}\sum_{V\in S_{N}}\frac{1}{N!}\sum_{W(\vec{n}) \in S_{N}}\exo{\left(\frac{2}{g^{2}}\Tr\left[W(\vec{n})\right]\right)}\frac{1}{N}\Tr\left[A V W(\vec{n})V^{\dag}B\right] \nonumber
\eea
\end{widetext}
where $A$ and $B$ represents the rest of the products of $W(\vec{n})$'s and we have used the fact that the action is invariant under $W(\vec{n})\rightarrow V W(\vec{n})V^{\dag}$ for arbitrary $V \in S_{N}$. For permutation matrices, it is possible to check that
\(
\label{normalization}
\frac{1}{N!}\sum_{V \in S_{N}} V_{ij}V_{kl}^{\dag}=\frac{1}{N}\delta_{il}\delta_{jk}+\frac{1}{N(N-1)}(J_{il}-\delta_{il})(J_{jk}-\delta_{jk})
\)
where $J$ is the  {\it Matrix-of-Ones} \cite{MoO}, a matrix with the number one in every slot. As a result of (\ref{normalization}), we can make the replacement $\Tr[AWB]\rightarrow (1/N)\Tr[W]\Tr[AB]$ to  leading order in $1/N$ in the expression above. Hence, the leading-$N$ term of the expectation value of Wilson loop operator can be written in the form
\(
W_{L}(g^{2},N)\approx [w(g^{2},N)]^{RT}
\)
up to subleading corrections in $N$ where 
\(
\label{wl for single plaq.}
w(g^{2},N)=\frac{1}{z}\frac{1}{N!}\sum_{W \in S_{N}}\exo{\left(\frac{2}{g^{2}}\Tr\left[W\right]\right)}\frac{1}{N}\Tr[W]
\)
This will be useful for us in evaluating the Wilson loop at large-$N$ when we discuss ``string tension".

\section{A First-Order Phase Transition at Large-$N$}

In analogy with the GWW example, we would like to identify an appropriate 't Hooft coupling to hold fixed, while sending $N$ to infinity.  Note that in vector, matrix and tensor models, the $N$-dependence of the respective 't Hooft couplings is distinct \cite{Klebanov}. One of our motivations for considering theories with alternate large-$N$ limits was the recent resurgence of interest in tensor models, see  \cite{ Klebanov, Witten1, CK1, CK2, CK3, CK4, CK5} for some points of entry into the literature. Since our example here is different from any of those (eg., the volume of the gauge group scales as $N!$ here as opposed to a power law), it is clear that the 't Hooft coupling must also be different. The latter $N$-dependence suggests that the entropy should scale as $\sim N \log N$, and together with the scaling of our action $\sim N/g^2$ we are lead to the guess that the 't Hooft coupling should scale as $g^2 \log N$. 

Note that the $N$-scaling for vector/matrix/tensor models are easily seen from their perturbative Feynman diagram expansion, whereas our arguments are from the lattice (and also implicitly from entropy, free energy, etc). It would be very interesting to connect these approaches. As far as we know, a Feynman diagram approach to gauge theories with finite groups has not been systematically undertaken in the literature. 

To make progress, let us change the index from $k_{1}$ to $j$ by writing $N-k_{1}=j$ and lift everything to the exponent in (\ref{subfactorial partition}):
\begin{align}
z(g^{2},N)&=\sum_{j=0}^{N} \exo{\left[\frac{2}{g^{2}}(N-j)+\ln{\frac{N!}{(N-j)!}}+\ln{\frac{!j}{j!}} \right]} 
\end{align}
The first term in the exponent is of $O(N)$ while the second term is of $O(N\ln{N})$ (from Stirling's approximation) in the large-$N$ limit. The only way in which these two terms can compete each other for a possible phase transition is if in the large-$N$ limit, $g^{2}\rightarrow 0$ in such a way that $\lambda \equiv g^{2}\ln{N}$ is a fixed quantity. We will see in the ensuing discussion that this choice indeed works as the appropriate choice of 't Hooft coupling.

Changing variables from $g^{2}$ to $\lambda$
\bea
z(\lambda,N)=\sum_{j=0}^{N} \exo\left[N\ln{N}\{S(j,\lambda, N)\}\right]
\eea
where the function $S(j,\lambda, N)$ is defined as:
\(
S(j,\lambda, N)\equiv\frac{2}{\lambda}(1-\frac{j}{N})+\frac{1}{N\ln{N}}(\ln{N!}-\ln{(N-j)!})+\ln{\frac{!j}{j!}}
\)
In the large-$N$ limit, $x\equiv j/N$ becomes a continuous variable. Writing the function $S(j,\lambda,N)$ in terms of $x$, we get
\begin{widetext}
\vspace{-0.2in}
\(
S(x,\lambda, N)=\frac{2}{\lambda}(1-x)+\frac{1}{N\ln{N}}(\ln{N!}-\ln{(N-Nx)!})+\ln{\frac{!(Nx)}{(Nx)!}}
\)
\end{widetext}
In the subsequent passages, we will see that the large-$N$ limit of this $S(x)$ function is related to the free energy of the system. An analogous free energy was defined in \cite{GW}. As a first step towards finding the large-$N$ behaviour of the $S(x)$ function, let us consider the asymptotic expansion of the subfactorial \cite{Asymp}:
\begin{widetext}
\bea
\label{subfactorial}
\frac{!(Nx)}{(Nx)!}=\frac{\sqrt{2\pi Nx}\l(\frac{Nx}{e}\r)^{N x}}{e (Nx)!}\l(1+\frac{1}{12Nx}+\frac{1}{288N^{2}x^{2}}+O\Big(\frac{1}{N^{3}x^{3}}\Big)\r)
+ \frac{(-1)^{Nx}}{Nx(Nx)!}\l(1-\frac{2}{Nx}+\frac{5}{N^{2}x^{2}}+O\Big(\frac{1}{N^{3}x^{3}}\Big)\r)
\eea
\end{widetext}
Using Stirling's approximation for large-$N$, $N!\sim \sqrt{2\pi N}(N/e)^{N}$, we see that the leading term in the first series (\ref{subfactorial}) is $e^{-1}\sim O(1)$ while the leading term in the second series is of $O(1/N^{N})$. Taking up to $O(1/N)$, in anticipation of the infinite $N$ limit, we can write: 
\begin{align}
    \ln{\frac{!(Nx)}{(Nx)!}} &\approx \ln\left[{e^{-1}\l(1+\frac{1}{12Nx}+O\l(\frac{1}{N^{2}}\r)\r)}\right] \nonumber \\
    &\approx -1 + \frac{1}{12Nx} + O\l(\frac{1}{N^{2}}\r)
\end{align}
Putting all these together, we write the $S(x)$ function as:
\begin{multline}
\label{S_x}
S(x,\lambda, N)\approx\frac{2}{\lambda}(1-x) + x - \frac{1}{\ln{N}}\l(x+(1-x)\ln{(1-x)}\r)\\
- \frac{1}{N\ln{N}}\l(\frac{1}{2}\ln{(1-x)}+1\r)+ O\l(\frac{1}{N^{2}\ln{N}}\r)
\end{multline}

\subsection{Leading-$N$}

The leading $N$ contribution to the integrand \eqref{S_x} in our case, is perhaps surprisingly, linear and therefore simple enough that one can explicitly do the integral! This is the first place where we see a hint that the physics in our case can have interesting differences from that seen in \cite{GW, Wadia1, Wadia2}. This key difference arises from the fact that both the action contribution and the measure contribution are simpler in our case, at leading-$N$. Doing the integral explicitly was not possible in \cite{GW, Wadia1, Wadia2} and one had to resort to a saddle point evaluation. As we will see shortly, while the integrand is simpler in our case, it is not so simple that we loose all interesting physics: we will still find a phase transition, albeit a first order one. 

The leading contribution to the partition function comes form the $O(1)$-terms in the $S(x)$-function:
\begin{align}
\label{leading partition}
    z(\lambda,N) &\approx \sum_{j} \exo\left[N\ln{N}\left\{\frac{2}{\lambda}(1-\frac{j}{N}) + \frac{j}{N}\right\}\right] \nonumber \\
    &= N \int_{0}^{1} dx\,\, \exo\left[N\ln{N}\left\{\l(1-\frac{2}{\lambda}\r)x +\frac{2}{\lambda} \right\}\right]\nonumber \\
    &= \frac{e^{N\ln{N}}-e^{\frac{2}{\lambda}N\ln{N}}}{\l(1-\frac{2}{\lambda}\r)\ln{N}}
\end{align}
An important physical quantity needed for our study of phase transition is the free energy defined by
\(
F(\lambda,N)\equiv -\frac{\lambda}{V}\ln{Z(\lambda,N)}=-\frac{\lambda}{a^{2}}\ln{z(\lambda,N)}
\)
We can build a scaled quantity from the free energy such that the scaled quantity has a well-defined large-$N$ limit:
\(
-E_{0}(\lambda,N)\equiv -\frac{Fa^{2}}{\lambda N\ln{N}}=\frac{\ln{z(\lambda,N)}}{N\ln{N}}
\)
This quantity is essentially the free energy per degrees of freedom. Let us examine the large-$N$ behaviour of $E_{0}$. The two terms in (\ref{leading partition}), $e^{N\ln{N}}$ and $e^{\frac{2}{\lambda}N\ln{N}}$, compete with each other and depending on the value of $\lambda$ one of them dominates the other. (It can be argued that the subleading-in-$N$ terms in the integrand \eqref{S_x} lead to subleading-in-$N$ terms in the integral. We will discuss the first leading correction in the next subsection.)  

In any event, in the large-$N$ limit the free energy as we have defined has a well-defined limit and it is captured by two different functions in the two different regimes: 
\begin{equation}
-E_{0}(\lambda) = \left\{
        \begin{array}{ll}
           \frac{2}{\lambda} & \quad \lambda \leq 2 \\
            1 & \quad  \lambda \geq 2
        \end{array}
    \right.
\end{equation}
Here, we use $E_0(\lambda)$ to denote  $E_0(\lambda, N \rightarrow \infty)$. 
The first-derivative of the free energy has a discontinuity at $\lambda=2$ and so we see that this phase transition is of first-order. 

\begin{figure}
\centering
\includegraphics[scale=0.5]{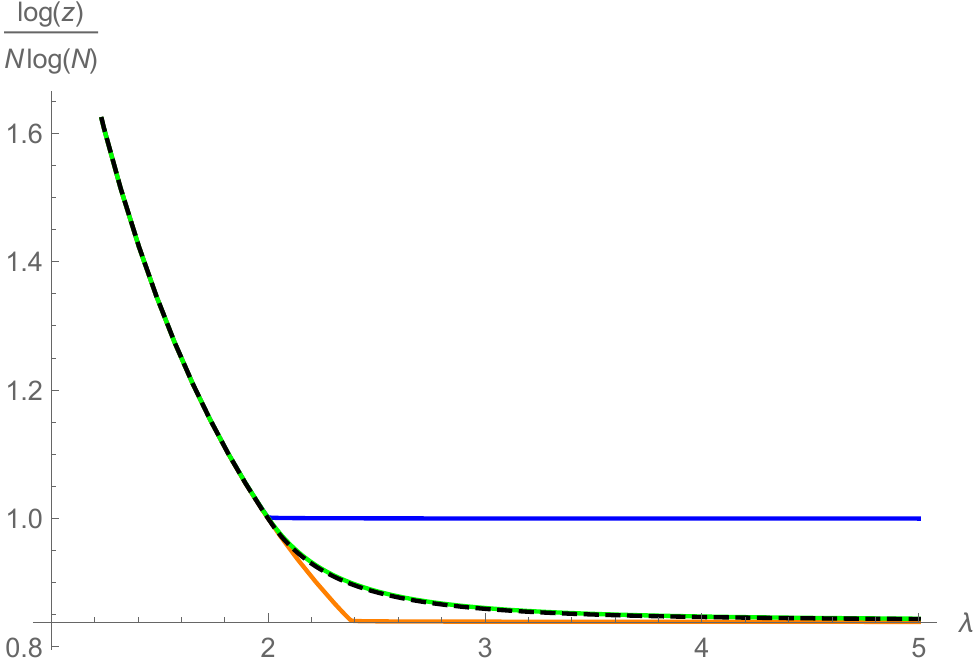}
\caption{The partition function at finite $N$, for $N=500$. Blue curve is the leading large-$N$ expectation, and shows the phase transition. The black curve is the exact partition function. For $\lambda< 2$ all curves overlap.  But for $\lambda >2$ the subleading correction is $O(1/\ln N)$, so the match at finite-$N$ is not perfect. The orange curve incorporates a crude analytically calculable piece of the $O(1/\ln N)$ correction, and shows that indeed the mismatch is of the correct order. The green dotted curve incorporates the exact (but numerical) $O(1/\ln N)$ correction, and it is visually indistinguishable in the plot from the exact result, the black curve. }
\label{plots for z}
\end{figure}

The above discussion is really enough to show the existence of the phase transition. But it turns out that the subleading fall-offs are fairly slow and go as $O(1/\ln N)$ in the $S_N$ theory as opposed to the power law fall-offs that are familiar from matrix or tensor models. So in plots at finite-$N$ for $N \lesssim 500$, the difference between the exact partition function and the asymptotic result above is noticeable.  So in the next subsection, we will compute these $O(1/\ln N)$ effects explicitly to check that indeed, these slow subleading fall-offs are precisely what are responsible  for this apparent mismatch. This will be viewed as evidence that in the strict large-$N$ limit, the existence of the phase transition is reliable.

\subsection{Subleading-$N$}

It is evident from Figure \ref{plots for z} that the leading contribution does not quite converge with the exact result even for $N$ as large as $500$. This slow convergence can be understood with the consideration of subleading corrections. The subleading terms of the function $S(x)$ in (\ref{S_x}) is of order $1/\ln N$:
\(
S(x,\lambda, N)\approx\frac{2}{\lambda}(1-x) + x - \frac{1}{\ln{N}}(x+(1-x)\ln (1-x))
\)

We can easily incorporate the piece $x/\ln N$ in the above expression, into our analytic treatment. This is instructive even though incomplete. Following the steps of the previous subsection with this term included, we arrive at a corrected form of $E_{0}$:
\begin{equation}
\label{1/ln N corrections}
-E_{0}\approx \left\{
        \begin{array}{ll}
           \frac{2}{\lambda} & \quad \lambda \leq 2 \\
            1-\frac{1}{\ln N} & \quad  \lambda \geq 2
        \end{array}
    \right.
\end{equation}
Plotting this at finite $N$ values, one can see that this correction is precisely the order of the mismatch between the exact and asymptotic results at finite-$N$, see Figure \ref{plots for z}.

But one can do better, and compute the exact subleading result. The partition function that includes the complete  $O(1/\ln N)$-correction can be expressed in the fairly elegant form:
\begin{widetext}
\(
z(\lambda,N)\approx N \int_{0}^{1} dx\,\, \exo\left[N\ln{N}\left\{\l(1-\frac{2}{\lambda}-\frac{1}{N}\r)x +\frac{2}{\lambda} \right\}\right] (1-x)^{-N(1-x)}
\)
\end{widetext}
We have not been able to do this integral analytically. However, the task can be done numerically very easily. The numerical result is depicted in Figure \ref{plots for z} along with our earlier analytical expressions. This precise incorporation of the $O(1/\ln N)$-term does not significantly alter the order of the correction, but it blunts the kink at $\lambda =2$, and we see that the numerical result with terms up to $O(1/\ln N)$ is a very good match to the exact result. The next lower order terms (of order $1/(N\ln N)$ and below) are essentially irrelevant. 

To summarize: the slow convergence to the leading order result can be attributed to the fact that the subleading corrections are of order $1/\ln N$. Of course, none of this changes the punchline about the existence of the phase transition at infinite $N$.

\section{Wilson Loop and ``String Tension"}

The expectation value of the Wilson loop operator can be easily computed by transforming the sum over group elements in (\ref{wl for single plaq.}) to a sum over the conjugacy classes (similar to what has been done for the partition function). With leading order terms in $S(x)$, we can write the Wilson loop expectation value for a single plaquette to be
\begin{widetext}
\bea
    w(\lambda,N)
    =\frac{1}{z(\lambda,N)}N\int_{0}^{1}dx\,(1-x)\exo\left[N\ln{N}\left\{\l(1-\frac{2}{\lambda}\r)x +\frac{2}{\lambda} \right\}\right] 
    =\frac{1}{1-e^{\l(1-\frac{2}{\lambda}\r)N\ln N}}+ \frac{1}{\l(1-\frac{2}{\lambda}\r)N\ln N}
\eea
\end{widetext}
The first term give the dominant contribution while the second term is subleading at order $1/(N\ln N)$. At large $N$, the behaviour of $w(\lambda,N)$ is roughly
\begin{equation}
w(\lambda, N\rightarrow \infty) \approx \left\{
        \begin{array}{ll}
           1+\frac{1}{\l(1-\frac{2}{\lambda}\r)N\ln N} & \quad \lambda \leq 2 \\
            \frac{1}{\l(1-\frac{2}{\lambda}\r)N\ln N} & \quad  \lambda \geq 2
        \end{array}
    \right.
\end{equation}
As in \cite{GW}, the Wilson loop expectation value  is seen to be less than unity in both regimes. This is a consequence of the fact that $\frac{1}{N}{\rm Tr}\ W \le 1$ for any $W$ simply because of the unitarity of $W$. In the strict $N$ going to infinity limit, it tends to unity in the weak coupling regime. This is what leads to the conclusion below that the ``string tension" we define vanishes at weak coupling.  

We can define a ``string tension"-like quantity, which is finite at infinite $N$:
\(
\sigma (N) \equiv -\frac{1}{a^{2}}\frac{\ln w(\lambda,N)}{\ln N}
\)
The re-scaling by $\ln N$ is necessary for us to get a finite result at large-$N$, and this is why we have put the ``string tension" in quotes. At large $N$ the ``string tension" becomes
\begin{equation}
\sigma\equiv \lim_{N\rightarrow \infty}\sigma(N) = \left\{
        \begin{array}{ll}
          0 & \quad \lambda \leq 2 \\
            1/a^2 & \quad  \lambda \geq 2
        \end{array}
    \right.
\end{equation}
In other words, this quantity jumps from zero to a finite value. Evidently, this is a very close analogue of the transition from a deconfining phase at weak coupling to a confining phase in the strong coupling regime.

\section{Comments and Outlook}

We have demonstrated that in the large-$N$ limit of $S_N$ gauge theory on a lattice, defined in a straightforward way following \cite{GW, Wadia1, Wadia2, Kogut}, there is a first order  phase transition from a deconfining-like phase to a confining-like phase as the coupling is cranked up. Even apart from the fact that we have found a large-$N$ transition in a theory with a tractable finite-$N$ partition function, this result is of some interest for a couple reasons. Firstly, we saw the emergence of a new type of 't Hooft coupling that was logarithmic in $N$ instead of power law as in matrix or tensor models. Secondly, the transition we saw, was a first order deconfinement-like transition. In the context of gauge-string duality, large-$N$ first order transitions on spheres have a long history \cite{Hawking, Witten}, see \cite{Shiraz, Dutta} for instructive discussions of various related matters. Despite the (superficial?) differences, it seems possible to us that deconfinement transition noted in \cite{Shiraz} may have connections to the transition we find here. Let us also note that the original GWW transition is a third order transition.

We have made some observations contrasting our set up and that in \cite{GW, Wadia1, Wadia2}, in the main text. Let us add some further comments along this direction here. In the GWW case, the phase transition is a result of the competition between the Haar measure and the tendency of a configuration to peak around the identity matrix. Here, the trade off between the dominant conjugacy class containing 1-cycles and the ``degeneracy" of the conjugacy class is a precise analogue. When one computes characters via traces as sums of eigenvalues, we are naturally doing the change of variables from the space of matrices to the space of their {\it sets of} eigenvalues. Unlike in the GWW case however, we find that in the strong coupling regime, the distribution of our eigenvalues is uniform for all $\lambda > 2$ and not just asymptotically as $\lambda \rightarrow \infty$. Finally, why is our phase transition first order as opposed to third order? In the $U(N)$ case, an understanding of the third order transition in terms of a Coulomb gas analogy was provided in \cite{Wadia2}. It may be useful to study how charged objects (say, spins on vertices) behave in our theory to get an idea of the mechanics of our phase transition -- right now we are only dealing with the ``pure glue" theory.

Our phase transition has both similarities and differences to the conventional phase transitions that one finds in lattice theories. These latter transitions are also deconfinement type transitions, but they are second order and they appear typically in higher than 1+1 dimensions \cite{Kogut}. Here on the other hand, we found a first order transition that is in 1+1 dimensions that is a result of the ``thermodynamic limit" arising from large-$N$.

As we mentioned in the introduction, our primary goal in considering non-Abelian finite groups as gauge groups, was the long-range goal of incorporating boundary gauge invariance into finite lattice models that are tractable as tensor network approaches to holography and bulk locality. This is clearly a direction worth further exploration. Closely related questions include: Can one add matter in the form of spins on vertices? What about higher dimensions? How are these models different from more familiar examples of discrete holographic toy models?

A further question that may be of interest here is the question of taking a continuum limit. Note that our gauge group is finite (and as a corollary, discrete), and continuum limits of such theories have not been widely studied in the literature. One exception is the paper \cite{Brezin} which studies a $\IZ_2$ gauge system coupled to matter without a formal continuum limit, and finds a (global) $\IZ_2$ invariant $\phi^4$ theory at a critical point. It may be interesting to study similar questions for non-Abelian finite groups like $S_N$. 

Finally, let us make one comment about other finite groups. Preliminary considerations suggest that suitable large-$N$ limits exist for other non-Abelian finite groups as well. It would be interesting to study this in detail. A similar comment applies to studying the $S_N$ theory with larger irreps -- note that exterior power representations of $S_N$ can become significantly larger than the standard representation for large values of $N$. Phase transitions in set ups of a loosely similar spirit were studied in a holographic context in \cite{Pallab1, Pallab2}. 

\section*{Acknowledgments}

We thank Prasad Hegde for helpful discussions on multiple lattice gauge theory related matters and comments on the draft. We also thank Spenta Wadia for detailed comments on the draft, and an entropy-based {\it a posteriori} argument for why our 't Hooft coupling is natural.    

\section*{APPENDIX A: $\IZ_{2}$, $\IZ_N$ AND $U(1)$}


The results we found in the main text appeared along with large-$N$ limits, non-Abelian gauge groups, and irreducible representations that scale in size with $N$. Let us illustrate how things fail when some of these assumptions are dropped by considering the case of the $\IZ_N$ gauge group which is Abelian (and has only 1- and 2-dimensional irreps).  
We start by following the methods of Sec. II to compute the partition function and Wilson loop for the $\IZ_{2}$ and $U(1)$ lattice gauge theories. 

For the $\IZ_{2}$ gauge group, the $W$ from (\ref{change variable}) now belongs to $\{+1,-1\}$ and consequently the partition function for a single plaquette (as written in (\ref{partition function})) becomes
\begin{align}
z&=\frac{1}{2}\sum_{W \in \{+1,-1\}}\exo{\left(\frac{1}{g^{2}}\Tr\left[W+W^{\dag}\right]\right)} \nonumber \\
&=\cosh{\l(2/g^{2}\r)}
\end{align}
The expectation value of the Wilson loop operator for a single plaquette in (\ref{wl for single plaq.}) can also be easily modified to
\begin{align}
    w&=\frac{1}{2z}\sum_{W \in \{+1,-1\}}\exo{\left(\frac{2}{g^{2}}\Tr\left[W\right]\right)}\frac{1}{2}\Tr[W] \nonumber \\
    &=\frac{1}{2}\tanh{\l(2/g^{2}\r)}
\end{align}

For the $U(1)$ gauge theory, in the one dimensional representation, we can define $W\equiv \exo{(i\theta)}$ where $\theta$ is a continuous variable belonging to the interval $[0,2\pi]$ (note that for the $\IZ_{2}$ case $\theta$ runs over only two values, $0$ and $\pi$). Changing variable from $W$ to $\theta$ the partition function corresponding to (\ref{partition function}) can be written as (the sum becomes an integral as $\theta$ is continuous):
\(
\label{u1 partition}
    z=\int_{0}^{2\pi} \frac{d\theta}{2\pi} \,\, \exo{\l(\frac{2}{g^{2}}\cos{\theta}\r)}=I_{0}(2/g^{2})
\)
where $I_{\alpha}$ is the modified Bessel functions of the first-kind for an arbitrary complex number $\alpha$. The generalization of this expression for $U(N)$ is the determinant of a Toeplitz matrix of Bessel functions \cite{Wadia2}. For the $U(1)$ theory, once again we compute the expectation value of the Wilson loop operator from (\ref{wl for single plaq.}) to be
\(
w=\frac{1}{z}\int_{0}^{2\pi}\frac{d\theta}{2\pi}\, e^{i\theta}\exo{\l(\frac{2}{g^{2}}\cos{\theta}\r)}=\frac{I_{1}(2/g^{2})}{I_{0}(2/g^{2})}
\)

Now we turn to the $\IZ_{N}$ Abelian gauge group. It is easy to see that it converges to the $U(1)$ theory in the large-$N$ limit. We take the one dimensional representation of $\IZ_{N}$: $W\equiv \exo{i(2\pi k/N)}$ where $k$ runs from $0$ to $N-1$. Hence, the normalized partition function for a single plaquette becomes
\(
\label{zn partition}
z = \frac{1}{N}\sum_{k=0}^{N-1}\exo{\l(\frac{2}{g^{2}}\cos{\l(\frac{2\pi k}{N}\r)}\r)}
\)
The above equation is analogous to the equation in the first line of (\ref{leading partition}) and we can proceed similarly by converting the sum to an integral. Making a change of variable from $k$ to $x(=k/N)$ (where $x$ becomes a continuous variable in the large-$N$ limit), we can write (\ref{zn partition}), for large-$N$, as
\begin{align}\label{zn partition function large n}
    z &= \int_{0}^{1} dx \,\,\exo{\l(\frac{2}{g^{2}}\cos{\l(2\pi x\r)}\r)} \nonumber\\
    &= \int_{0}^{2\pi} \frac{d\theta}{2\pi} \,\, \exo{\l(\frac{2}{g^{2}}\cos{\theta}\r)}
\end{align}
where in the second equality there is one more change of variable: $x$ to $\theta(=2\pi x)$. The expression in (\ref{zn partition function large n}) is identical to (\ref{u1 partition}) and hence we explicitly established that the large-$N$ limit of $\IZ_{N}$ is actually $U(1)$ theory, as expected.

The above formulas where they overlap with \cite{Kogut}, agree with the results there.  Moreover, it has been shown in \cite{Kogut} that the string tension is finite for both the weak and strong coupling regime. Hence, the Abelian gauge theory in 1+1 dimension is always in confinement and there is no phase transition. Of course, in our $S_N$ case we are able to  get past this because of the qualitative differences brought about by the infinite number of degrees of freedom at large-$N$. Let us also clarify a potentially confusing point (at least it was, for us) about Abelian theories on the lattice. The UV completions of these theories are provided by the lattice and not by any field theory, and there is no continuum limit in the UV (unlike say, in lattice QCD). When the statement that $U(1)$ gauge theory is the continuum limit is made (see eg., \cite{Kogut}), it is to be understood that one is referring to the deep IR where the lattice spacing has become irrelevant and we can trust the field theory description.   

Finally, let us also comment that one can get past some of these objections by considering large but {\it reducible} representations. But since this has the spirit of working with many copies of same (or similar) objects and seems less intrinsic a property of the gauge group under question, we will not undertake it here. 


\end{document}